\documentclass[english,twocolumn,superscriptaddress,aps,pra]{revtex4-1}
\usepackage[T1]{fontenc}
\usepackage[latin9]{inputenc}
\usepackage{color}
\usepackage{amsmath}
\usepackage{graphicx}

\makeatletter

\@ifundefined{textcolor}{}
{%
\definecolor{BLACK}{gray}{0}
\definecolor{WHITE}{gray}{1}
\definecolor{RED}{rgb}{1,0,0}
\definecolor{GREEN}{rgb}{0,1,0}
\definecolor{BLUE}{rgb}{0,0,1}
\definecolor{CYAN}{cmyk}{1,0,0,0}
\definecolor{MAGENTA}{cmyk}{0,1,0,0}
\definecolor{YELLOW}{cmyk}{0,0,1,0}
}

\makeatother

\usepackage{babel}
\begin{document}

\title{Detecting faked continuous variable entanglement using one-sided
device-independent entanglement witnesses }

\author{B. Opanchuk, L. Arnaud and M. D. Reid}

\affiliation{Centre for Quantum and Optical Science, Swinburne University of Technology,
Melbourne, 3122 Australia}
\begin{abstract}
We demonstrate the principle of one-sided device-independent continuous
variable (CV) quantum information. In situations of no trust, we
show by enactment how the use of standard CV entanglement criteria
can mislead Charlie into thinking Alice and Bob share entanglement,
when the data is actually generated classically using a Local Hidden
Variable theory based on the Wigner function. We distinguish between
criteria that demonstrate CV entanglement, and criteria that demonstrate
the CV Einstein-Podolsky-Rosen (EPR) steering paradox. We show that
the latter, but not the former, are necessarily one-sided device-independent
entanglement witnesses, and can be used by Charlie to signify genuine
EPR entanglement, if he trusts only Alice. A monogamy result for the
EPR steering paradox confirms security of the shared amplitude values
in that case.
\end{abstract}
\maketitle

\section{Introduction}

The importance of \emph{trust}, either of an observer or of the devices
employed, has motivated the emergence of ``a new paradigm''~\cite{gallego}
in the field of quantum information: device-independent (DVI) quantum
information processing~\cite{acinbell,acinbell2,bancal ent di,bancalnew,guh4,gallego,liupanletetr}.
In light of this emergence, it is now recognised that quantum nonlocality
has a very special role to play in quantum information. In a classic
scenario, Charlie wants to confirm whether the qubit values reported
to him by Alice and Bob have been generated from an entangled quantum
source. For quantum key distribution (QKD), this may allow him to
deduce that these values are not known to any eavesdropper, Eve~\cite{cry scheme benn,bellcry}.
Quite surprisingly, if Charlie uses certain entanglement criteria,
called device-independent entanglement witnesses (DVIEW), he does
not have to assume anything about the exact nature of the devices
Alice and Bob are using or the reliability of their measurements~\cite{acinbell,acinbell2,bellmon2,bellcry5}.

Device-independent entanglement witnesses are those introduced by
Bell, that demonstrate not only quantum entanglement, but quantum
nonlocality~\cite{acinbell,acinbell2,bancal ent di}. If Charlie
verifies violation of a Bell inequality, he can be sure the statistics
did not arise from \emph{any} local hidden variable (LHV) theory~\cite{Bell}.
The concept of quantum nonlocality is distinct from that of entanglement,
and DVI security is not given by all entangled states. A problem
for the implementation of DVI quantum information however is that
when most photons are not detected, not all LHV theories can be eliminated~\cite{norbert,LS,bellloophole,limgisinloop,loopbellfake}.
These sorts of ``loophole'' issues are often overcome by continuous
variable (CV) quantum information. Here, information is carried in
the amplitudes of fields, enabling efficient detection and transmission~\cite{cv gauss,expcv,cvmarscommu,recentcvtele,sca,hanncvone,earlcvcryepr}.
A second  problem then arises. The most widely-used CV entangled
resource (the two-mode squeezed state) has a LHV description, for
the results of Alice and Bob's amplitude measurements~\cite{ou,rmp-1}.
This means that those amplitude values will not violate a Bell inequality.

The amplitude values of the CV entangled resource\emph{ }do display
quantum nonlocality however: namely, the nonlocality presented by
Einstein-Podolsky-Rosen (EPR)~\cite{epr} in their 1935 EPR paradox~\cite{mdrepr,mdeprprl}.
EPR's nonlocality addresses whether the action of measurement by Bob
can immediately influence the results of Alice's \emph{quantum} measurements
at another distant site. EPR demonstrate their paradox, by assuming
that values at Alice's\emph{ }location arise from measurements of
quantum observables. There is no similar assumption for Bob's measurements.
The EPR paradox is realised through a procedure that is ``one-sided
device-independent'', and criteria for the EPR paradox, different
to those for ``EPR entanglement'', are ``one-sided'' DVIEW\@.

The concept of ``\emph{one-sided device-independent}'' security
has been pioneered in Refs.~\cite{hw2-steering-1,hw-steering-1,steer ent cry},
based on the resource of quantum steering~\cite{Schrodinger}. In
fact, the EPR paradox is an example of quantum steering~\cite{hw-steering-1,EPRsteering-1,hw2-steering-1}.
The analysis~\cite{steer ent cry} focused on QKD with qubits, but
revealed advantages, if one is justified in assuming trust at \emph{one}
site. The advantages are implicit in previous work on CV QKD based
on criteria that relate to the EPR paradox~\cite{earlcvcryepr,hanncvone,mdrin ficek}.
Clearly, this concept has the potential to introduce the ``new DVI
paradigm'' to CV quantum information.

Here, we demonstrate the principle and feasibility of CV one-sided
DVI quantum information. We do this in three steps: (1) First, we
illustrate the vulnerability of CV quantum information in situations
of no trust, if one uses standard entanglement criteria such as that
derived by Duan et al~\cite{Duan-simon-1-1} and Simon~\cite{simon-1-1}.
Our illustration is by enactment: where neither Alice and Bob can
be trusted, we deceive Charlie by simulating entanglement trivially
using a classical computer. We next show that even if Alice can be
trusted, we can still deceive Charlie if he uses a criterion that
is not a one-sided DVIEW\@. The ``faked'' entanglement can be generated
using an unsophisticated hybrid scheme involving classical computers
and a classical optical source. We also discuss the motivation by
a malicious Eve to fake the entanglement $-$ so that she can fool
Charlie into thinking a set of amplitude values are held securely
between the two parties, Alice and Bob, when in fact they can be distributed
by her to an infinite number of parties.

(2) Second, we prove that if Charlie confirms a one-sided DVI EPR
paradox criterion \emph{and} he trusts Alice, he can be sure the results
\emph{do }originate from an EPR entangled state.

(3) Lastly, we use a monogamy result~\cite{monogamy,mdrin ficek,monog2}
for the EPR paradox to show that confirmation of the EPR paradox criterion
is enough to give security against an eavesdropper Eve possessing
a noise-free replica of the amplitude values, \emph{despite} that
Bob cannot be trusted.

 Practical implementation of CV one-sided DV security requires demonstration
of an EPR-steering paradox, which is more generally difficult than
entanglement~\cite{eprrobust}. Nonetheless, the CV EPR paradox has
been verified without fair sampling loopholes for a range of optical
systems~\cite{ou,rmp-1,recenthannepr}. These will be briefly discussed
in the Conclusion. The elimination of the ``locality'' loophole
is also important~\cite{loopbellfake}, if CV DVI is to become a
reality. Since security is often comprised at one site only, we expect
CV one-sided DVI protocols to become useful.

\section{Detecting Shared Entanglement}

Let Alice and Bob be two spatially separated observers at stations
labelled $A$ and $B$ respectively. Suppose an observer at a central
station sends to Alice one of the quantum subsystems that comprise
a CV EPR state, and to Bob the other. A sequence of similarly prepared
states is transmitted. After each transmission $t_{i}$, an observer
Charlie selects randomly to ask Alice and Bob to measure either the
position $X$ or the momentum $P$ of the subsystems at their respective
stations (or sites), and to report to him the values of their measurements~\cite{cry scheme benn,bellcry,sca,hanncvone,earlcvcryepr,cvmarscommu}.
A CV EPR state is a simultaneous eigenstate of momentum-sum and position-difference,
and the values given by Alice and Bob will be accordingly correlated.
If Charlie determines that these values satisfy an entanglement criterion,
then he can confirm that Alice and Bob indeed share an entangled state.

\subsection{Duan-Simon criteria for CV entanglement}

The most widely-used CV EPR entanglement criterion is based on the
methods of Duan et al and Simon~\cite{simon-1-1,Duan-simon-1-1,proof for product form}.
Entanglement between the systems measured by Alice and Bob is confirmed
if
\begin{equation}
\mathrm{Ent}=\frac{4}{(1+g^{2})}\Delta(X_{A}-gX_{B})\Delta(P_{A}+gP_{B})<1\label{eq:eny}
\end{equation}
where $g$ is any real constant and we use the notation $(\Delta X)^{2}\equiv\langle X^{2}\rangle-\langle X\rangle^{2}$.
Here, $X_{A/B}$ and $P_{A/B}$ are the ``position'' and ``momentum'
quadratures measured by Alice and Bob respectively, and we have selected
a suitable scaling so that the Heisenberg quantum uncertainty principle
is written $\Delta X\Delta P\geq1/4$. Clearly, the CV EPR state will
satisfy the condition~\eqref{eq:eny}, with $g=1$.  For more general
states, the choice of $g$ is taken optimally so that the left-side
of the inequality (\ref{eq:eny}) is minimised. For the Gaussian systems
that are most commonly utilised for CV QIP \cite{cv gauss}, and restricting
to the subclass of Gaussian states with symmetry between position
and momentum correlation, this criterion can be shown to be \emph{necessary
and sufficient} for detecting bipartite entanglement \cite{buono}.

Before continuing, it is important to understand the assumptions needed
in deriving the criterion (\ref{eq:eny}) (and other very similar
criteria), since it is these assumptions that create the security
loopholes addressed by device-independent quantum information processing.
To derive the criterion (\ref{eq:eny}), one begins by supposing the
bipartite system of Alice and Bob is non-entangled. By definition
of entanglement, this means that the bipartite density operator can
be written in the separable form \cite{Ent}
\begin{equation}
\rho_{AB}=\sum_{R}P(R)\rho_{A}^{R}\rho_{B}^{R}\label{eq:sep}
\end{equation}
where here $\sum_{R}P(R)=1$ and $\rho_{A}^{R}$ and $\rho_{B}^{R}$
are quantum density operators for Alice's system $A$ and Bob's system
$B$ alone. $ $The uncertainty product for the separable mixture
is constrained by the uncertainty product for the product states that
are the components of the mixture. For product states $\rho_{A}^{R}\rho_{B}^{R}$
we can write
\begin{eqnarray}
\Delta_{R}(X_{A}-gX_{B})\Delta_{R}(P_{A}+gP_{B})\,\,\,\,\,\,\,\,\,\,\nonumber \\
\geq\Delta_{R}X_{A}\Delta_{R}P_{A}+g^{2}\Delta_{R}X_{B}\Delta_{R}P_{B}\label{eq:proofstep1}
\end{eqnarray}
where we write the subscript $R$ to remind use the averages are with
respect to the component state $R$ (see Refs. \cite{proof for product form,Duan-simon-1-1,simon-1-1}
for full details). Using the fact that $\rho_{A}^{R}$ and $\rho_{B}^{R}$
are \emph{quantum} states, the Heisenberg uncertainty relation applies
to each, and it follows that
\begin{eqnarray}
\Delta_{R}(X_{A}-gX_{B})\Delta_{R}(P_{A}+gP_{B}) & \geq & \frac{1}{4}(1+g^{2})\label{eq:proofstep2}
\end{eqnarray}
Thus, if (\ref{eq:eny}) is satisfied, the two systems $A$ and $B$
cannot be represented by (\ref{eq:sep}) and must therefore be entangled.

The crucial assumption for the practical application of the criterion
by Charlie is that \emph{both} sets of values, $X_{A}$, $P_{A}$
and $X_{B}$, $P_{B}$, are truly outcomes of measurements of the
quantum observables, and are hence correctly constrained for all possible
local component states $\rho_{A}^{R}$, $\rho_{B}^{R}$ by the Heisenberg
uncertainty relations: $\Delta_{R}X_{A}\Delta_{R}P_{A}\geq1/4$ and
$\Delta_{R}X_{B}\Delta_{R}P_{B}\geq1/4$. If Alice and Bob do not
report the results of quantum measurements, the criterion cannot be
applied to faithfully detect entanglement \cite{bancal ent di,hw-steering-1,steer ent cry}.
\emph{Both} \emph{Alice and Bob must be trusted}.

We remark that the well-known simple ``Duan'' criterion \cite{Duan-simon-1-1}
detects entanglement between $A$ and $B$ if
\begin{equation}
\bigl(\Delta(X_{A}-X_{B})\bigr)^{2}+\bigl(\Delta(P_{A}+P_{B})\bigr)^{2}<1\label{eq:duansimple}
\end{equation}
 This criterion follows directly from the criterion (\ref{eq:eny}),
by taking $g=1$ and noting that for any real numbers $x$ and $y$,
$x^{2}+y^{2}\geq2xy$. Hence, if the simple criterion detects entanglement,
so will the more powerful criterion (\ref{eq:eny}) \cite{proof for product form},
provided $g$ is optimally chosen. Thus, the results of this paper
remain valid if criterion (\ref{eq:duansimple}) is used instead of
(\ref{eq:eny}).

\subsection{EPR-steering criterion as a one-sided DVI criterion for entanglement}

Entanglement can also be confirmed by the EPR paradox~\cite{ou,epr}.
Such a paradox (also called an EPR steering paradox~\cite{hw-steering-1,hw2-steering-1,EPRsteering-1})
is realised if
\begin{equation}
\mathrm{EPR}_{A|B}=4\Delta_{\mathrm{inf}}X_{A|B}\Delta_{\mathrm{inf}}P_{A|B}<1\label{eq:eprsteer-2}
\end{equation}
where $(\Delta_{\mathrm{inf}}X_{A|B})^{2}$ is the variance of the
conditional distribution for the measurement $X_{A}$ given a measurement
at $B$~\cite{mdrepr}, and $(\Delta_{\mathrm{inf}}P_{A})^{2}$ is
defined similarly. The subscript ``inf'' reminds us of the original
formulation of the EPR paradox, in which the emphasis is on an observer,
here called Bob, who can make a near-perfect inference of the outcome
of a measurement by Alice, even though he is spatially separated from
Alice's measurement station. For Gaussian systems, meaning Gaussian
states and measurements, this criterion has been shown \emph{necessary
and sufficient} to detect EPR steering \cite{hw2-steering-1}.

On selecting $(\Delta_{\mathrm{inf}}X_{A})^{2}$ and $(\Delta_{\mathrm{inf}}P_{A})^{2}$
to be the variances conditional on measurements $X_{B}$, and $P_{A}$,
respectively, we see that the CV EPR state will satisfy this EPR condition.
The EPR paradox criterion can then more specifically be written as
\begin{equation}
\mathrm{EPR_{A|B}}=4\Delta(X_{A}-gX_{B})\Delta(P_{A}+gP_{B})<1\label{eq:eprsteer-1}
\end{equation}
where $g$ is a real constant optimised so that the left side of inequality
(\ref{eq:eprsteer-1}) is minimised.

We now briefly summarise the assumptions needed to derive the EPR
criteria, for comparison with those needed to derive the Duan/ Simon-type
entanglement criteria. The key point is that the EPR steering criteria
are based on fewer assumptions \cite{steer ent cry}, and are useful
for CV one-sided DVI quantum information processing.

The EPR criteria can be derived using the extension of EPR argument,
in which Bob is able to predict the result for Alice's $X_{A}$ or
$P_{A}$ measurement to an uncertainty given by $\Delta_{inf}X$ and
$\Delta_{inf}P$, respectively \cite{mdrepr,rmp-1}. In their argument,
EPR make the assumption that certain premises (referred to as EPR's
local realism LR) will hold. They use these premises to deduce that
there is a local realistic description for Alice's measurements, in
which the results for $X_{A}$ and $P_{A}$ are \emph{simultaneously}
predetermined to a precision $\Delta_{inf}X$ and $\Delta_{inf}P$,
which of course when $\Delta_{inf}X_{A}\Delta_{inf}P_{A}<1/4$ contradicts
any local quantum state description. Then, the EPR paradox is established,
because LR implies a local description at $A$ that is inconsistent
with (the completeness of) quantum mechanics.

The alternative proof in which the criterion is proven to be one for
EPR-steering involves similar assumptions \cite{EPRsteering-1}. In
fact, the proofs for EPR steering follow along parallel lines to those
for entanglement. For the EPR-steering proof, separability takes the
more general form of Bell's Local Hidden Variable (LHV) model \cite{Bell}.
This may be likened to first step in the EPR argument, where the assumption
of EPR's local realism is made \cite{mdrin ficek}. On expansion,
we see that the criteria involve second order moments such as $\langle X_{A}^{\theta}X_{B}^{\phi}\rangle$
where $X_{A/B}^{\theta}=X_{A/B}\cos\theta+P_{A/B}\sin\theta$ ($\theta=0$
or $\pi/2$). For all LHV models, these can be expressed in the separable
form
\begin{equation}
\langle X_{A}^{\theta}X_{B}^{\phi}\rangle=\int P(\lambda)\langle X_{A}^{\theta}\rangle_{\lambda}\langle X_{B}^{\phi}\rangle_{\lambda}d\lambda\label{eq:LHVmodel-1}
\end{equation}
where $\lambda$ are classical ``hidden variable'' parameters and
$P(\lambda)$ is the probability distribution for these parameters.
The $\lambda$ correspond to the $R$ in (\ref{eq:sep}) and at the
various steps in the proof. The $\langle X_{A}^{\theta}\rangle_{\lambda}$
is the average value for the result $X_{A}^{\theta}$, given the hidden
variable state specified by $\lambda$. The $\langle X_{B}^{\phi}\rangle_{\lambda}$
is defined similarly. Manipulation based on this LHV assumption reveals
the result (\ref{eq:proofstep1}) will once more hold, but with $\lambda$
replacing $R$. However, after this step, different to the proof for
entanglement, the two subsystems are treated asymmetrically. For Alice's
measurements, consistency with local quantum states $\rho_{A}^{\lambda}$
is again assumed, so that the Heisenberg uncertainty relation holds
i.e. $\Delta_{\lambda}X_{A}\Delta_{\lambda}P_{A}\geq\frac{1}{4}$.
This may be likened to the step in the EPR argument where the ``elements
of reality'' for Alice's local states are compared for consistency
with quantum mechanics \cite{mdrin ficek}. However, no such assumption
is made about Bob's local system or measurements. It is only assumed
that the variances are positive i.e. $\Delta_{\lambda}X_{B}\Delta_{\lambda}P_{B}\geq0$.
With these assumptions, we find
\begin{eqnarray}
\Delta_{\lambda}(X_{A}-gX_{B})\Delta_{\lambda}(P_{A}+gP_{B}) & \geq & \frac{1}{4}\label{eq:proofstep2-1}
\end{eqnarray}
The LHV model with the additional constraint for local states at $A$
is called an LHS model \cite{hw-steering-1}. This LHS model will
imply $\Delta(X_{A}-gX_{B})\Delta(P_{A}+gP_{B})\geq\frac{1}{4}$.
Violation of the inequality (\ref{eq:eprsteer-1}) is therefore an
EPR paradox, and quantum steering of $A$ by $B$ \cite{hw-steering-1,EPRsteering-1}.

We see from both derivations that it is necessary to assume Alice
does indeed report the results of the appropriate quantum measurements,
if the EPR steering criterion is to be valid. However, different to
the Duan-Simon entanglement criteria, \emph{i}t is not necessary to
make this assumption about Bob's measurements. Thus, the criterion
is independent of the devices used at Bob's measurement station. Regardless
of the orgin of Bob's values, the criterion is still valid for detecting
the EPR steering paradox. \emph{Only Alice must be trusted}.

In detecting the EPR steering paradox, the criterion will \emph{also}
detect entanglement \cite{rmp-1,hw-steering-1}. This is because any
separable description (\ref{eq:sep}) will also imply the LHV expansion
(\ref{eq:LHVmodel-1}), with or without the assumption of the uncertainty
relation $\Delta_{\lambda}X_{A}\Delta_{\lambda}P_{A}\geq\frac{1}{4}$,
and therefore cannot generate an EPR steering paradox. Separable models
are subsets of LHV and LHS models. The EPR criterion is therefore
a \emph{one-sided device-independent criterion }for entanglement.

It has been observed~\cite{eprrobust} that the EPR paradox criterion~\eqref{eq:eprsteer-2}
to less robust to noise and losses than the Duan-Simon entanglement
criterion~\eqref{eq:eny}. The EPR paradox criterion does not detect
all entangled states. This is because of the different assumptions
made about the nature of Bob's measurements. For Eq.~\eqref{eq:eny},
the measurements $X_{B}$ and $P_{B}$ at Bob's location must genuinely
correspond to those of the quantum conjugate observables. For the
EPR criterion, this assumption about Bob's measurements is\emph{ }not\emph{
}made. The EPR criterion~\eqref{eq:eprsteer-2} makes fewer assumptions
than can be justified if one can indeed trust Alice and Bob to make
the right measurements, and is as a consequence less sensitive to
detecting entanglement. The criterion is more generally difficult
to satisfy, but has the advantage of being (equivalent to) a one-sided
DVIEW~\cite{bancal ent di,acinbell,acinbell2}.

\begin{figure}
\begin{centering}
\includegraphics[width=0.8\columnwidth]{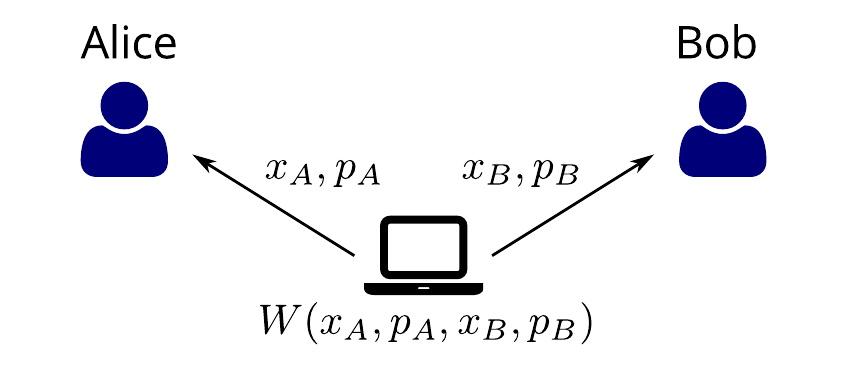}
\par\end{centering}

\caption{\emph{\label{fig:Faking-the-CV}Faking the CV entanglement using classical
protocols where the measurement stations of Alice and Bob cannot be
trusted:} Two classical number pairs representing position and momentum
are generated probabilistically (by Eve) from the computer using the
Wigner function \eqref{eq:wig} and are sent to Alice and Bob. Alice
and Bob fake entanglement by reporting these numbers to Charlie, who
then verifies ``entanglement'' using either the Duan-Simon criterion~\eqref{eq:eny}
or the EPR criterion~\eqref{eq:eprsteer-2} (Figure 3, curves (a)
and (b)). Security is compromised because the amplitude values can
be shared by an infinite number of parties in this case. }
\end{figure}

\section{Faking CV entanglement}

It may be that Alice and Bob (or their devices) \emph{cannot} be trusted,
and do not report to Charlie the values of measured quantum quadrature
phase amplitudes. Instead, they may report values that have been created
by a classical computer simulation. Then, we will show that the entanglement
criteria of the last section can be satisfied even though there is
no entanglement. We call this ``faked entanglement''. The faking
of the entanglement is possible, because the assumptions made in the
derivation of the criteria are no longer valid. In this paper, we
illustrate the faking of entanglement in two scenarios: first, where
neither Alice nor Bob can be trusted (Figure 1), in which case neither
criterion \eqref{eq:eny} or \eqref{eq:eprsteer-2} is valid; and
second, where only Alice can be trusted (Figure 2), in which case
only the EPR criterion \eqref{eq:eprsteer-2} is valid.

The implications can be understood by considering a motivation for
the faking protocol. In the scenario of Section II, Charlie has transmitted
to Alice and Bob the particles or light fields that form the two subsystems
of an entangled EPR state. He wants Alice and Bob to report the results
of their measurements of either $X$ or $P$ at each site, for a sequence
$t_{i}$ of such transmissions, so that he can confirm entanglement
by taking the appropriate averages required by the entanglement criteria
(\ref{eq:eny}) or (\ref{eq:eprsteer-2}). Once he confirms entanglement,
the amplitude values that are shared by Alice and Bob become useful
for QIP tasks, for example, for CV quantum key distribution (QKD)
\cite{cry scheme benn,sca}. The sequence $t_{i}$ that is used to
test for entanglement does not have to be exhaustive, so a sequence
of unrevealed correlated amplitudes can remain secret to the stations
of Alice and Bob. Now suppose a devious Eve seeks to trick Charlie,
Alice and Bob. We will show how she can create from a computer a sequence
of numbers $x_{A},$ $p_{A}$, $x_{B}$, $p_{B}$ that mimic the results
for measurements $X_{A},$ $P_{A},$ $X_{B}$, $P_{B}$ respectively,
in that they will satisfy the inequalities of the entanglement criteria
\eqref{eq:eny} and \eqref{eq:eprsteer-2}. We call these ``Eve's
numbers''. Each pair $x_{A}$, $p_{A}$ of the sequence can be stored,
in order, at Alice's station, and each associated pair $x_{B}$, $p_{B}$
can be stored, in order, at Bob's station. Alice and Bob may be accomplices
of Eve, or else it may be they are loyal to Charlie, but that their
devices (e.g. a final readout computer) are tampered with by an infiltrator.
Either way, if Charlie trusts the results that Alice and Bob report
and believes they hold a sequence of values for amplitudes based on
an entangled state, there could be serious consequences. The amplitudes
of the whole sequence are \emph{not} secure, but can be shared by
an infinite number of parties, since the sequence is formerly known
to Eve and can be distributed by her. The validity of QKD based on
the assumption of an entangled state is destroyed.

We now illustrate by example the method of faking entanglement. The
standard realisation of the CV EPR state is the two-mode squeezed
state~\cite{tmss}:
\begin{equation}
|\psi\rangle={\rm sech}(r)\sum_{n=0}^{\infty}\tanh^{n}(r)|n\rangle_{A}|n\rangle_{B}\label{eq:tmss}
\end{equation}
 where $|n\rangle_{A/B}$ are the number states of modes at the locations
$A/B$ of Alice and Bob respectively, and $r\geq0$ is the squeeze
parameter that determines the amount of entanglement between the modes
of Alice and Bob. The Wigner function for this state is
\begin{eqnarray}
W(\underline{x}) & = & \frac{4}{\pi^{2}}\exp\{-[(x_{A}-x_{B})^{2}+(p_{A}+p_{B})^{2}]/\sigma_{-}^{2}\nonumber \\
 &  & -[(x_{A}+x_{B})^{2}+(p_{A}-p_{B})^{2}]/\sigma_{+}^{2}\}\label{eq:wig}
\end{eqnarray}
where $\sigma_{+}^{2}=\exp(2r)$ and $\sigma_{-}^{2}=\exp(-2r)$ and
we introduce the notation $\underline{x}=(x_{A},p_{A},x_{B},p_{B})$~\cite{wigner}.
The moments of the quadrature phase amplitudes are given directly
by the amplitude moments of the distribution. The positivity of the
Wigner function allows for a perfectly accurate probabilistic simulation
of the entangled \emph{quadrature} correlations, based on the measurements
of $X_{A/B}^{\theta}$. This fact, pointed out by Bell \cite{bell book},
is well-known \cite{ou}. The implication is that if the Charlie restricts
to a criterion based on quadrature measurements, and if Alice and
Bob or their devices cannot be trusted, he can be fooled into thinking
there is entanglement shared between them when there is not.

Hence, an alternative realisation of the process of detecting the
correlations given by~\eqref{eq:eny} or~\eqref{eq:eprsteer-2}
is as follows. At transmission time $t_{i}$, a central source Eve
generates with probability $W(\underline{x})$ the values $x_{A}$,
$p_{A}$, $x_{B}$ and $p_{B}$, and sends them to Alice and Bob (Figure~\ref{fig:Faking-the-CV}).
The process is repeated $N$ times, a different set of numbers being
probabilistically generated on each trial $t_{i}$. When Charlie calls
for the value of transmission $t_{i}$, Alice gives to him the value
$x_{A}$ or $p_{A}$, depending on whether Charlie asked her (and
Bob) to measure position or momentum. Similarly, Bob reports his value,
$x_{B}$ or $p_{B}$. Charlie evaluates the averages as required for
the criteria~\eqref{eq:eny} and~\eqref{eq:eprsteer-2}. The averages
are precisely consistent with the predictions of the two-mode squeezed
state, and both criteria are satisifed for all $r$. Specifically,
we find (on selecting the optimal value for $g$ in (\ref{eq:eny}),
namely $g=1$)
\begin{eqnarray}
\mathrm{Ent} & = & \exp(-2r)\label{eq:sol-1}
\end{eqnarray}
and (on optimising the choice of measurements at $B$ and selecting
the optimal value for $g$ in (\ref{eq:eprsteer-1}), namely $g=\tanh2r$
\cite{mdrepr})
\begin{eqnarray}
\mathrm{EPR}_{A|B} & = & {\rm sech}(2r)\label{eq:eprsoln}
\end{eqnarray}

We carry out the procedure of Figure~\ref{fig:Faking-the-CV}, and
illustrate the faking of the quantum entanglement by classical means
in Figures~\ref{fig:observables-1000},~\ref{fig:observables-100}
and~\ref{fig:wigner-points}. A possible set of ``Eve's numbers''
is presented in Figure \ref{fig:wigner-points}. Despite that the
data satisfy the entanglement criteria (\ref{eq:eny}), (\ref{eq:duansimple})
and (\ref{eq:eprsteer-2}), (\ref{eq:eprsteer-1}), there is \emph{no
possibility of security}, because the amplitude values can be distributed
from the source to an infinite number of other parties. The undermining
of the security occurs because neither of the entanglement criteria~\eqref{eq:eny}
or~\eqref{eq:eprsteer-2} used by Charlie are (equivalent to) DVIEW.
The particular moments evaluated by Charlie can be constructed using
a Local Hidden Variable (LHV) theory~\cite{acinbell,acinbell2} based
on the use of the positive Wigner function (\ref{eq:wig}) \cite{ou,bell book}.
This is true for all Gaussian states, defined as describable by a
positive Gaussian Wigner function \cite{cv gauss}. As explained in
the last Section, the second order moments needed for the criteria~\eqref{eq:eny}
and~\eqref{eq:eprsteer-2} are $\langle X_{A}^{\theta}X_{B}^{\phi}\rangle$
where $X_{A/B}^{\theta}=X_{A/B}\cos\theta+P_{A/B}\sin\theta$ ($\theta=0$
or $\pi/2$). For LHV models, these can be written as
\begin{equation}
\langle X_{A}^{\theta}X_{B}^{\phi}\rangle=\int P(\lambda)\langle X_{A}^{\theta}\rangle_{\lambda}\langle X_{B}^{\phi}\rangle_{\lambda}d\lambda\label{eq:LHVmodel}
\end{equation}
where $\lambda$ are classical ``hidden variable'' parameters and
$P(\lambda)$ is the probability distribution for these parameters.
The $P(\lambda)$ is independent of the choice of measurement ($\theta$
and $\phi$), which is made \emph{after} the generation and separation
of the subsystems. The $\langle X_{A}^{\theta}\rangle_{\lambda}$
is the average value for the result $X_{A}^{\theta}$, given the hidden
variable state specified by $\lambda$, and the $\langle X_{B}^{\phi}\rangle_{\lambda}$
is defined similarly. In the classical simulation of Figures~\ref{fig:Faking-the-CV},~\ref{fig:observables-1000}
and~\ref{fig:observables-100}, the Wigner variables $\underline{x}$
assume the role of Bell's hidden variables $\lambda$, $W(\underline{x})$
becomes $P(\lambda)$, and there is a deterministic situation whereby
$\langle X_{A}^{\theta}\rangle_{\lambda}=x_{A}^{\theta}\equiv x_{A}\cos\theta+p_{A}\sin\theta$
and $\langle X_{B}^{\phi}\rangle_{\lambda}=x_{B}^{\phi}\equiv x_{B}\cos\phi+p_{B}\sin\phi$,
so that $\langle X_{A}^{\theta}X_{B}^{\phi}\rangle=\langle x_{A}^{\theta}x_{B}^{\phi}\rangle$
\cite{bell book,ou}.

A DVIEW would, like a Bell inequality, negate all LHV models of type
(\ref{eq:LHVmodel}). Thus use of an DVIEW entanglement criterion
would allow Charlie to detect entanglement, by eliminating all of
the LHV models, and therefore represents a more secure test of entanglement
\cite{acinbell,acinbell2}. However, for the CV entanglement generated
by the two-mode squeezed state, a DVIEW cannot be constructed directly
from the amplitude $X$ and $P$ measurements only, since the LHV
model (\ref{eq:LHVmodel}) describes the statistics perfectly in this
case \cite{bell book,ou}. We point out that, in view of the work
of Banaszek and Wodkiewicz \cite{PRA wignerbell}, this does not exclude
DVI entanglement witnesses being constructed from other sorts of measurements.
Also, the violation of a Bell inequality has been shown possible using
quadrature phase amplitudes measurements, for more complex EPR-type
sources \cite{bell violation}.

\section{Two trusted measurement stations }

It should be emphasised that Charlie can safely conclude genuine entanglement
based on the criteria~\eqref{eq:eny} and~\eqref{eq:eprsteer-2},\emph{
}if\emph{ }he can trust that the ``measurement values'' reported
by Alice and Bob are genuinely the results of the quantum observables
$\hat{X}_{A/B}$ and $\hat{P}_{A/B}$.  Let us re-examine the derivation
of the criteria, as outlined in Section II, to clarify this point.

Entanglement is defined as the failure of \emph{quantum }separability.
For any quantum separable state~\cite{Ent}, $\rho_{AB}=\sum_{R}P_{R}\rho_{A}^{R}\rho_{B}^{R}$,
which implies the moments can be expressed as  (on identifying $R$
as a hidden variable $\lambda$)
\begin{equation}
\langle X_{A}^{\theta}X_{B}^{\phi}\rangle\equiv\langle\hat{X}_{A}^{\theta}\hat{X}_{B}^{\phi}\rangle=\int P(\lambda)\langle x_{A}^{\theta}\rangle_{\lambda}\langle x_{B}^{\phi}\rangle_{\lambda}d\lambda\label{eq:m3}
\end{equation}
where the $\langle x_{A}^{\theta}\rangle_{\lambda}$ and $\langle x_{B}^{\phi}\rangle_{\lambda}$
are \emph{additionally} constrained to be the moments of \emph{quantum}
states $\rho_{A}^{\lambda}$ and $\rho_{B}^{\lambda}$. Each predetermined
local state (as specified by $\lambda$) is required to satisfy the
quantum uncertainty relations. It is this fact which prevents a perfect
correlation between Alice and Bob's ``real'' positions and momenta
$-$ unless there is entanglement. The criteria~\eqref{eq:eny} and~\eqref{eq:eprsteer-2}
falsify the quantum separable model~\eqref{eq:m3}, but not the local
hidden variable model~\eqref{eq:LHVmodel}.

\begin{figure}
\begin{centering}
\includegraphics[width=0.8\columnwidth]{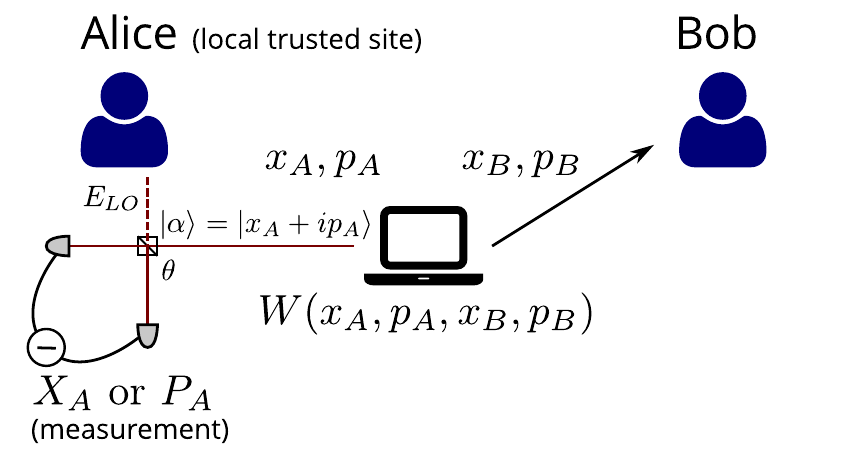}
\par\end{centering}

\caption{\emph{\label{fig:Faking-the-CV-1}Faking the CV entanglement in the
case where Alice can be trusted: }The scenario is similar to Figure
1 but now Alice can be trusted to report to Charlie the results of
measurements of genuine quantum observables $\hat{X}_{A}$ and $\hat{P}_{A}$.
Bob however cannot be trusted. Entanglement satisfying the Duan-Simon
criterion~\eqref{eq:eny} can still be faked using the asymmetric
hybrid protocol depicted, involving the computer and a coherent state
light source $|\alpha\rangle$ (where $\alpha=x_{A}+ip_{A}$) that
is deviously inputed to Alice's site (Figure~\ref{fig:observables-1000},
curve (c)). The entanglement detected by the EPR criterion~\eqref{eq:eprsteer-2}
however cannot be faked by one-sided scenarios where Alice is trusted
(Figure 3, curve (d)). If violated in the one-sided scenario, the
EPR criterion detects genuine entanglement, and (through the monogamy
result~\eqref{eq:monog}) gives security against all processes that
would allow distribution of the amplitude values. In the picture,
Alice measures the quadrature amplitudes $\hat{X}_{A}$ and $\hat{P}_{A}$
of the field at her location using a standard homodyne measurement
scheme, whereby the field is combined across a beam splitter with
an intense local oscillator field (denoted $E_{LO}$). The variable
phase shift $\theta$ allows her to choose whether $\hat{X}_{A}$
or $\hat{P}_{A}$ is measured. }
\end{figure}

In the classical simulation based on (\ref{eq:LHVmodel}), the predetermined
local states are given by the classical numbers $\{x_{A},p_{A}\},\{x_{B},p_{B}\}$,
all of which are predetermined with perfect accuracy: That is, they
are \emph{not} constrained by the uncertainty relation. If Alice
and Bob's detectors have been tampered with or are unreliable, the
values reported could arise from such a classical simulation in which
case there can be an unlimited number of replica sets $\{x_{B},p_{B}\}$
distributed.

\begin{figure}
\begin{centering}
\includegraphics{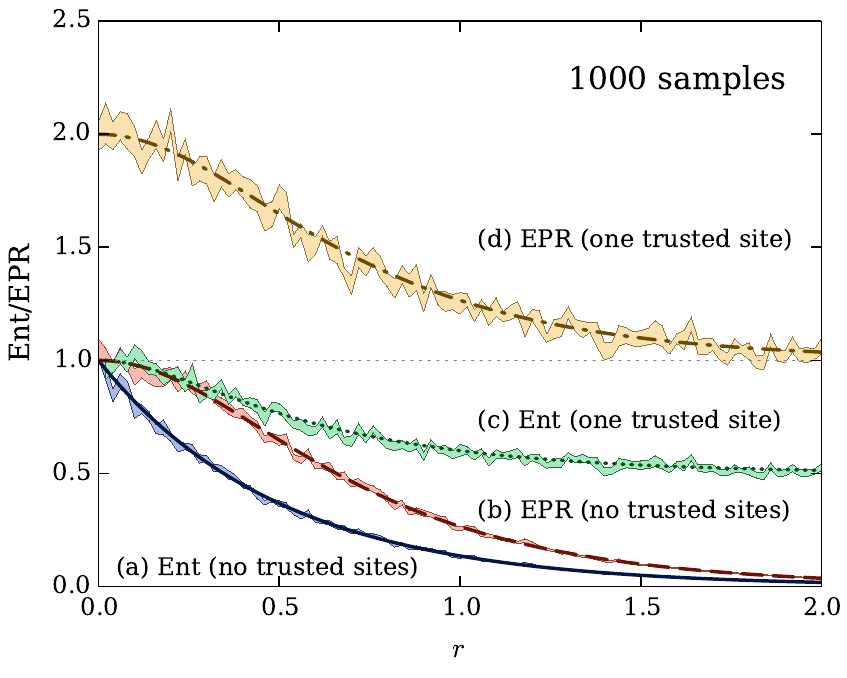}
\par\end{centering}

\caption{\emph{\label{fig:observables-1000}Distinguishing faked from genuine
entanglement using a one-sided device-independent entanglement criterion:
} Curves (a) and (b) plot the results for $Ent$ and $EPR_{A|B}$
as realised by the enactment of the classical protocol of Figure~\ref{fig:Faking-the-CV},
where neither Alice nor Bob can be trusted. Here, $r$ is the squeeze
parameter used in the Wigner function and there are $N=1000$ trials.
The bandwidths correspond to the sampling errors, centred on the sampled
means. Both the entanglement certified by the Duan-Simon and that
certified by the EPR steering criterion are faked. Charlie can be
fooled into thinking there is entanglement, if he mistakingly trusts
Alice and Bob. Curves (c) and (d) plot the results for the enactment
of the classical hybrid protocol of Figure~\ref{fig:Faking-the-CV-1},
where Alice can be trusted. The Duan-Simon entanglement criterion~\eqref{eq:eny}
can be faked (curve c), but \emph{not} the EPR steering inequality~\eqref{eq:eprsteer-2}
(curve d), which is a one-sided device-independent entanglement criterion.
If Charlie trusts Alice, he can use the EPR criterion to distinguish
faked from genuine entanglement.}
\end{figure}

\begin{figure}
\begin{centering}
\includegraphics{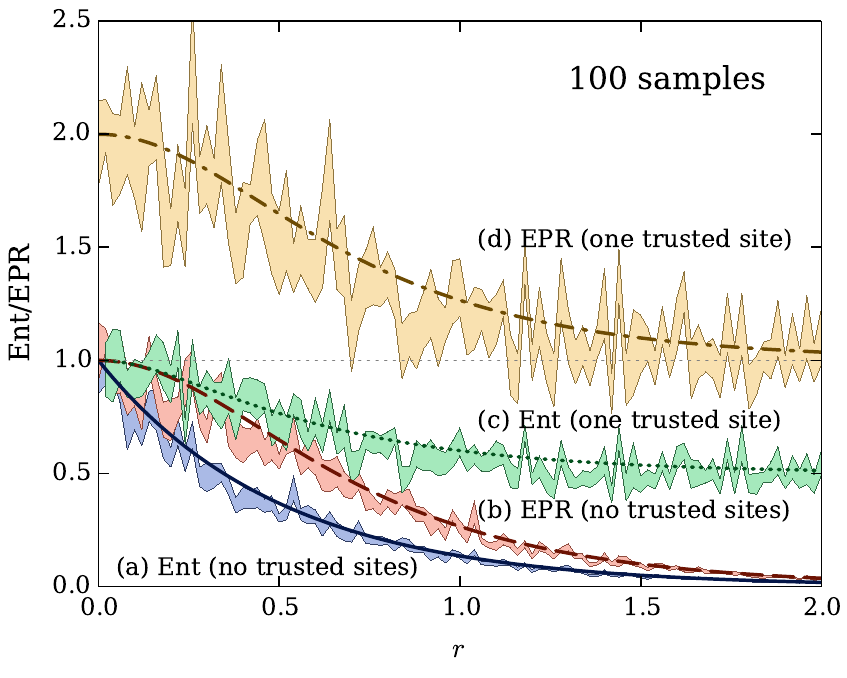}
\par\end{centering}

\caption{\emph{\label{fig:observables-100}Distinguishing faked from genuine
entanglement using a one-sided device-independent entanglement criterion:
} Curves as for Figure~\ref{fig:observables-1000}, but where $N=100$. }
\end{figure}

\begin{figure}
\includegraphics{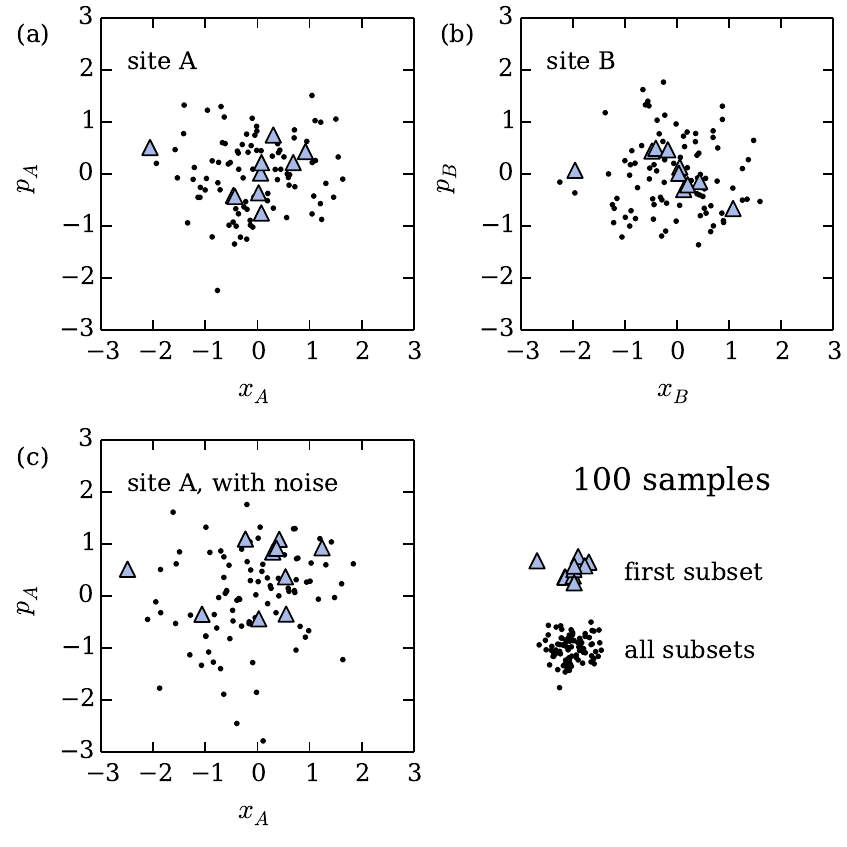}

\caption{\label{fig:wigner-points}\emph{Actual values reported by Alice and
Bob:} (a) and (b) give a possible set of $\underbar{x}=(x_{A},p_{A},x_{B},p_{B})$
for $N=100$ trials where $r=0.7$. Here, pane (a) gives the set of
values for Alice, and pane (b) those for Bob. These are ``Eve's numbers'',
used for the purpose of faking entanglement (Figure \ref{fig:Faking-the-CV}).
The pane (c) shows values for Alice's where the results are generated
by the hybrid protocol of Figure~\ref{fig:Faking-the-CV-1}, in which
case the values $x_{A}$ and $p_{A}$ correspond to the results of
genuine quantum measurements $\hat{X_{A}}$ and $\hat{P_{A}}$. In
this case, noise is added to the data. Triangles correspond to the
first subset of 10 samples used to calculate one of the samples for
$\mathrm{EPR_{A|B}}$ or $\mathrm{Ent}$.}
\end{figure}

\section{One trusted measurement station}

We now turn to the case where \emph{one} measurement station, that
of Alice, can be trusted. In this scenario, we will show that the
use of the one-sided DVIEW criterion~\eqref{eq:eprsteer-2} enables
reliable witnessing of entanglement: That is, faked entanglement is
detected as ``not-entangled'', and the criterion is only satisifed
when there is genuine entanglement. We expect the one-sided entanglement
witness to be useful, since protocols are often asymmetric with one
station being secure than the other \cite{hw-steering-1,steer ent cry}.

Any classical LHV model~\eqref{eq:LHVmodel} becomes constrained,
if the moments generated from it are to be consistent with statistics
arising from this new scenario. The moments are given by a separable
model
\begin{equation}
\langle X_{A}^{\theta}X_{B}^{\phi}\rangle\equiv\langle\hat{X}_{A}^{\theta}X_{B}\rangle=\int P(\lambda)\langle x_{A}^{\theta}\rangle_{\lambda}\langle p_{B}^{\phi}\rangle_{\lambda}d\lambda\label{eq:m4}
\end{equation}
where \emph{only} $\langle x_{A}^{\theta}\rangle_{\lambda}$ is constrained
to be consistent with a quantum density operator $\rho_{A}^{\lambda}$~\cite{hw-steering-1}.
Unlike for the model~\eqref{eq:m3}, there is \emph{no} similar constraint
on $B$. This hybrid separable model, which is a superset of all quantum
separable models~\eqref{eq:m3}, is called a Local Hidden State (LHS)
model. The falsification of this model indicates the nonlocality of
``quantum steering''~\cite{hw-steering-1,hw2-steering-1}.

It has been proved in Ref.~\cite{EPRsteering-1} that the EPR inequality
$\mathrm{EPR}_{A|B}\geq1$ holds for all LHS models~\eqref{eq:m4}.
We have outlined this proof in Section II. Hence, the EPR criterion~\eqref{eq:eprsteer-2}
(which shows an ``EPR steering'' of Alice's system) negates all
of the classical LHS models~\eqref{eq:m4}. In this way, entanglement
is demonstrated, \emph{without} the assumption of quantum measurements
for the values reported at Bob's location. On the other hand, the
LHS model~\eqref{eq:m4} does \emph{not} imply the entanglement criterion~\eqref{eq:eny}.
In short, statistics generated by a LHS model~\eqref{eq:m4} \emph{can}
satisfy the entanglement criterion~\eqref{eq:eny}, but \emph{cannot}
satisfy the EPR paradox criterion~\eqref{eq:eprsteer-2}.

We give a demonstration (Figure~\ref{fig:Faking-the-CV-1}) of how
Charlie can be mislead into thinking there is entanglement, if he
uses the wrong criterion for this scenario whereby only Alice can
be trusted. Bob in collusion with Eve may set up the following protocol.
Alice would be expecting (at each trial or time $t_{i}$ of the sequence)
an input to her station, corresponding to the subsystem of the entangled
state sent by Charlie. This subsystem may take the form of an optical
beam or pulse, the mode corresponding to $A$ in the two-mode squeezed
state (\ref{eq:tmss}). Her station is secure, so Eve and Bob cannot
tamper with it. Instead, they decide to tamper with the input. Inputed
to Alice's station at each trial are the values $x_{A}$, $p_{A}$
that stem from the central classical computer source, as before. However,
these cannot be the final inputs, as Alice passes on to Charlie only
the results of genuine quantum measurements $\hat{X}_{A}$, $\hat{P}_{A}$,
made locally on a real experimental system. Bob then comes up with
a hybrid protocol: Alice's experimental system is deviously coupled
to the input values, so that the measurements she reports are those
of $\hat{X}_{A}$, $\hat{P}_{A}$ on the quantum coherent state $|\alpha\rangle$
where $\alpha=x_{A}+ip_{A}$. For example, the optical beam or pulse
that arrives at Alice's station is not the subsystem $A$ of the entangled
state (\ref{eq:tmss}), but the optical coherent state $|\alpha\rangle$.
The statistics from this hybrid classical scenario are describable
by an LHS model~\eqref{eq:m4}. Although the extra local quantum
uncertainty provided by the coherent state means that Bob and Eve
cannot predict exactly the values that Alice will report, the correlation
with the values $x_{B}$, $p_{B}$ (which they do know) is good enough
that the inequality of the Duan-Simon criterion (\ref{eq:eny}) is
satisfied (though not maximally). This makes it possible to a fake
entanglement certified by the Duan-Simon criterion.

The results of carrying out Bob and Eve''s protocol are given in Figure~\ref{fig:observables-1000}.
In our simulation, Alice does not actually measure the amplitudes
of the optical coherent state. Instead, we model the outcomes of her
local measurement using standard quantum theory. We note that the
simulation results are in agreement with the predictions of a two-mode
squeezed state~\eqref{eq:tmss} which is then coupled at Alice's
location to a local quantum noise source. The final mode for Alice's
values is $X_{A'}=X_{A}+X_{\mathrm{coh}}$ where $\langle X_{\mathrm{coh}},X_{\mathrm{coh}}\rangle=1/4$.
Calculation gives
\begin{eqnarray}
\mathrm{Ent} & = & \frac{4}{(1+g^{2})}\{n-2gc+g^{2}m\}\label{eq:sol-2}
\end{eqnarray}
where $n=\langle X_{A'},X_{A'}\rangle=\frac{1}{4}\{1+\cosh(2r)\}$,
$m=\langle X_{B},X_{B}\rangle=\cosh(2r)/4$ and $c=\langle X_{A}X_{B}\rangle=\sinh(2r)/4$,
and $\langle X,Y\rangle=\langle XY\rangle-\langle X\rangle\langle Y\rangle$.
The entanglement measured by~\eqref{eq:eny} is optimised for the
choice $g=(n-m+\sqrt{(n-m)^{2}+4c^{2}})/(2c)$. For Gaussian distributions,
the minimum possible $[\Delta_{\mathrm{inf}}(X_{A})]^{2}$ is given
by the minimum value of $\langle[\Delta(X_{A'}-g_{\mathrm{epr}}X_{B})]^{2}\rangle=n+g_{\mathrm{epr}}^{2}m-2g_{\mathrm{epr}}c$
where $g_{\mathrm{epr}}$ is a real constant~\cite{rmp-1}. Selecting
$g_{\mathrm{epr}}=c/m$, we find
\begin{eqnarray}
\mathrm{EPR}_{A|B} & = & 1+{\rm sech}(2r)
\end{eqnarray}
which, consistent with our knowledge of the LHS model~\eqref{eq:m4},
\emph{cannot} satisfy~\eqref{eq:eprsteer-2}. Yet, we see (Figure~\eqref{fig:observables-1000})
that the entanglement criterion~\eqref{eq:eny} is satisfied for
all values of $r$.

We remark that all of the ``EPR steering'' inequalities~\cite{EPRsteering-1,eprsteer,eprsteerphoton}
that falsify the model~\eqref{eq:m4} will suffice as one-sided DVIEW~\cite{steer ent cry}.
One such criterion is \cite{ferrisbec,rmp-1}
\begin{equation}
\mathrm{Ent}<0.5\label{eq:otherduanhalfcrit}
\end{equation}
as evident in Figure~\eqref{fig:observables-1000}. However, for
Gaussian states and measurements, the EPR criterion~\eqref{eq:eprsteer-2}
has been shown necessary and sufficient, whereas the criterion (\ref{eq:otherduanhalfcrit})
is not \cite{hw-steering-1,hw2-steering-1}. The EPR criterion is
therefore generally more sensitive, as has been shown useful when
asymmetry exists between the stations of Alice and Bob \cite{onewaysteer,thermal epr,steer thmurray,kate2}.
The entropic EPR steering criteria of Ref.~\cite{eprsteer} are likely
to be useful in more complex scenarios.

In summary, if Charlie establishes that the statistics reported to
him give an EPR paradox $\mathrm{EPR}_{A|B}<1$ \emph{and} he trusts
Alice's values, then he can be sure that there is genuine entanglement
between the two observers. If his data falls in the region of entanglement
but with no EPR paradox, he cannot be sure of this. The statistics
could be consistent with a classical protocol, and the values $x_{A}$,
$p_{A}$ known to an infinite number of parties (Figure~\ref{fig:Faking-the-CV-1}).

\section{Security and the Monogamy of the EPR paradox\emph{ }}

In confirming a genuine EPR paradox, Charlie can deduce the security
of the amplitude values, based on a knowledge of quantum mechanics.
As with Bell's nonlocality~\cite{bell mon,bellmon2,bellmongisin},
the EPR paradox satisfies a very strict form of monogamy. It is always
true that (unless quantum mechanics fails or we cannot trust Alice's
devices)
\begin{equation}
\mathrm{EPR}_{A|B}\mathrm{EPR}_{A|E}\geq1\label{eq:monog}
\end{equation}
where $E$ is any other system measured by an eavesdropper, Eve~\cite{monogamy,mdrin ficek,monog2}.
The monogamy relation (\ref{eq:monog}) is derived based on the assumption
of the uncertainty principle for $\hat{X}_{A}$, $\hat{P}_{A}$, and
is one-sided DVI, that is, is valid no matter what devices or measurements
are used by Bob, or Eve. Similar to the relations introduced in Refs.~\cite{uncerqubitcry,steer ent cry},
the mere measurement of $\mathrm{EPR}_{A|B}$ allows Charlie to deduce
the level of security of the correlation between Alice and Bob's amplitudes.
If $\mathrm{EPR}_{A|B}<1$ then $\mathrm{EPR}_{A|E}>1$, and the conditional
variances that give the noise levels on Eve's inferences of Alice's
amplitudes is increased by an amount known to Charlie.

\section{Discussion and Conclusion\emph{ }}

Here, we have demonstrated the principle of one-sided device-independent
quantum security as applied to continuous variables. We believe this
principle could expedite the practical application of device-independent
cryptography and may lead to more secure protocols in CV quantum information.

We have shown that one-sided device-independent security is possible,
using the widely available two-mode squeezed state CV EPR source and
quadrature phase amplitude measurements. The detection of entanglement
using the EPR steering criterion is well documented in the literature,
and has been achieved without detection loophooles (though not as
yet without locality loopholes). This quadrature amplitude EPR steering
has been realised in a number of different optical scenarios \cite{ou,bow,fr,eprrobust,buono},
most of which are explained in the review of Ref. \cite{rmp-1}. Also,
more recently, very high degrees of EPR steering correlation have
been reported \cite{recenthannepr}. Significant for application to
quantum communication protocols is that the CV EPR steering criterion
has been verified for pulses propagating through optical fibres \cite{fiber},
and has now been confirmed for spatially entangled optical modes and
networks \cite{wag,boy,fabtreps}.

The biggest drawback to any practical implementation is the relative
lack of robustness of the EPR paradox criterion to losses. It has
been shown theoretically using monogamy relations that the criterion
cannot be satisfied when there is 50\% or more loss of the intensity
received by the steering (untrusted) party \cite{monog2}. For losses
up to 50\% on this channel, however, the criterion has been experimentally
verified \cite{buono,eprrobust}. On the other hand, the criterion
is quite robust with respect to the losses on the trusted party \cite{eprrobust,steer thmurray,onewaysteer}.
As detection can be done very efficiently using homodyne techniques,
the greatest problem is likely to be the losses that enter on transmission.
This may make the choice of location of the trusted and untrusted
measurement stations important, as has been discussed in relation
to applications of photonic steering \cite{steer ent cry,teleportation stations}.
All this gives promise however to the prospect of utilising CV one-sided
device-independent witnesses for quantum information tasks.

It is stressed that two-sided device-independent entanglement witnesses
could also be applied to CV scenarios. However, these may require
more complicated measurements and protocols.
\begin{acknowledgments}
This work was supported by the Australian Research Council Discovery
Projects scheme. MDR thanks the invitation to Ringberg Castle.\end{acknowledgments}

\end{document}